\documentclass[fleqn,usenatbib,useAMS]{mnras}

\usepackage{graphicx}	
\usepackage{amsmath}	
\usepackage{amssymb}	
\usepackage{multicol}        
\usepackage{multirow}
\usepackage{bm}		
\usepackage{pdflscape}	
\usepackage{subfigure}
\usepackage{color}

\usepackage[T1]{fontenc}
\usepackage{ae,aecompl}

\usepackage{newtxtext,newtxmath}

\title[Low-frequency GMRT observations of ultra-cool dwarfs]{Low-frequency GMRT observations of ultra-cool dwarfs}

\author[A. Zic et al.]{Andrew Zic,$^{1,2}$\thanks{E-mail: \href{mailto:azic7771@uni.sydney.edu.au}{azic7771@uni.sydney.edu.au}}
Christene Lynch,$^{1,2}$
Tara Murphy,$^{1,2}$
David L. Kaplan,$^{3}$
Poonam Chandra$^{4}$\\
\\
$^1$ Sydney Institute for Astronomy, School of Physics, The University of Sydney, NSW 2006, Australia\\
$^2$ ARC Centre of Excellence for All-sky Astrophysics (CAASTRO)\\
$^3$ Department of Physics, University of Wisconsin--Milwaukee, Milwaukee, WI 53201, USA\\
$^4$ National Centre for Radio Astrophysics, TIFR, Pune University Campus, Pune 411007, India}
\date{Accepted XXX. Received YYY; in original form ZZZ}

\pubyear{2018}

\begin{document}
\label{firstpage}
\pagerange{\pageref{firstpage}--\pageref{lastpage}}
\maketitle

\begin{abstract}
Observations of radio emission in about 10 per cent of ultra-cool dwarfs (UCDs) indicate the presence of strong, persistent magnetic fields in these stars. These results are in contrast to early theoretical expectations on fully-convective dynamos, and to other tracers of magnetic activity, such as $\mathrm{H}\,\alpha$ and X-ray luminosity. Radio-frequency observations have been key to physically characterising UCD magnetospheres, although explaining the diverse behaviour within them remains challenging. Most radio-frequency studies of UCDs have been conducted in the 4-8~GHz band, where traditional radio interferometers are typically most sensitive. Hence, the nature of UCD radio emission at low frequencies ($\lesssim 1.4~\text{GHz}$) remains relatively unexplored, but can probe optically thick emission, and regions of lower magnetic field strengths -- regimes not accessible to higher-frequency observations. In this work, we present the results from Giant Metrewave Radio Telescope observations of nine UCDs taken at $\sim610$ and $1300~\text{MHz}$. These are the first observations of UCDs in this frequency range to be published in the literature. Using these observations, we are able to constrain the coronal magnetic field strength and electron number density of one of the targets to $1 \lesssim B \lesssim 90~\mathrm{G}$ and $4 \lesssim \log(N_e) \lesssim 10$, respectively. We do not detect the flaring emission observed at higher frequencies, to a limit of a few millijanskys. These results show that some UCDs can produce low-frequency radio emission, and highlights the need for simultaneous multi-wavelength radio observations to tightly constrain the coronal and magnetospheric properties of these stars.
\end{abstract}






\section{Introduction}
Stellar magnetic fields influence stars through all evolutionary stages, affecting their formation, mass loss, rotation, chemical composition, and behaviour at late evolutionary stages \citep{2009ARA&A..47..333D}.
At masses below $0.3-0.4~\text{M}_\odot$ (spectral types later than about $\text{M}3.5$), stellar interiors are fully convective \citep{2000ARA&A..38..337C}, and lack the tachocline generally accepted to drive solar-type dynamos \citep{ 2003A&ARv..11..287O}. Therefore, early expectations were that these stars were incapable of generating and sustaining strong, large-scale magnetic fields \citep{1993SoPh..145..207D}. 

However, flares observed in very late M-dwarfs in $\mathrm{H}\,\alpha$, UV, and X-ray \citep{1990A&A...231..131T, 1995ApJ...455..670L, 1999MNRAS.302...59M, 1999ApJ...519..345L, 1999ApJ...527L.105R, 2000ApJ...533..372F} provided indications that some very low-mass stars exhibit at least sporadic magnetic activity. 
More recent work (e.g. \citealp{2017MNRAS.465.1995M,2016MNRAS.457.1224L,2015ApJ...799..192W,2015ApJ...815...64W,2010ApJ...709..332B,2008A&A...487..317A,2008ApJ...684..644H,2006ApJ...653..690H,2002ApJ...572..503B,2001Natur.410..338B}) on ultra-cool dwarfs (UCDs; spectral class later than $\sim$M7) at radio frequencies have revealed that around $10\%$ of these stars exhibit both persistent and bursty radio emission, with many violating the well-known G\"udel-Benz relation (GBR) by several orders of magnitude \citep{2014ApJ...785....9W}. The GBR is an empirically derived relation between X-ray and radio luminosities of magnetically active FGKM stars spanning several orders of magnitude, following $L_X \sim L_{\nu,R}^{0.73}$ \citep{1994A&A...285..621B}, with $L_X$ denoting X-ray luminosity and $L_{\nu,R}$ radio spectral luminosity. 

Gyrosynchrotron emission from a population of mildly-relativistic electrons (e.g. \citealt{2005ApJ...627..960B}, see \citealt{1985ARA&A..23..169D} for a review), or the electron-cyclotron maser instability (ECMI; e.g \citealt{2008ApJ...684..644H}, see \citealt{2006A&ARv..13..229T} for a review) have both been argued to be responsible for UCD radio emission. The ECMI is generally thought to be responsible for the bursty, flare-like emission from UCDs, owing to brightness temperatures in excess of $10^{12}~\mathrm{K}$, and up to 100 per cent circularly polarised emission. On the other hand, the persistent, non-flaring emission from UCDs has been argued to originate from gyrosynchrotron (e.g. \citealp{2017MNRAS.465.1995M, 2016MNRAS.457.1224L, 2015ApJ...815...64W, 2005ApJ...627..960B}) or ECMI (e.g. \citealp{2018ApJ...854....7L, 2014ApJ...783...27S,2008ApJ...684..644H}) processes.  In any case, the presence of this radio emission reveals that these objects can generate strong, axisymmetric and active magnetic fields,
in contrast to the expectations set by early fully-convective dynamo theory \citep{1993SoPh..145..207D}.
Despite recent progress in dynamo models of low-mass stars 
(e.g. \citealt{2016ApJ...833L..28Y, 2015ApJ...813L..31Y, 2010SSRv..152..565C, 2008ApJ...676.1262B}), explaining the broad phenomenology of UCD magnetic activity revealed through their radio emission remains challenging \citep{2014ApJ...785....9W}.

The level of magnetic activity revealed by the radio emission was also surprising, given steep declines observed for UCDs in X-ray and $\mathrm{H}\,\alpha$ luminosity (e.g. \citealp{2014ApJ...785....9W, 2008ApJ...673.1080B, 2004AJ....128..426W, 2003ApJ...583..451M, 2000AJ....120.1085G}), which are conventional tracers of coronal and chromospheric activity in FGKM stars.
It is thought that the drop in $\mathrm{H}\,\alpha$ and X-ray luminosities is indicative of the transition to cooler, more neutral atmospheres for UCDs \citep{2002ApJ...571..469M}. The declines in $\mathrm{H}\,\alpha$ and X-rays are accompanied by a breakdown in the rotation-activity relation for UCDs (e.g. \citealp{ 2003ApJ...583..451M}), while the radio rotation-activity relation remains relatively constant in the UCD regime \citep{2012ApJ...746...23M}. These divergent trends once again highlight the need to clarify UCD dynamo and coronal properties.

Radio-frequency observations have proven to be the most effective probes of UCD magnetism. This is in part due to the  difficulty encountered when attempting techniques such as Zeeman--Doppler imaging to the faint, rapidly-rotating UCDs. This technique has only recently been successfully demonstrated for a UCD \citep{2017ApJ...847...61B}.

However, the majority of radio studies of UCDs have been conducted between 4 and 8~GHz, and low-frequency ($\lesssim 1.4~\text{GHz}$) radio emission of UCDs remains relatively unexplored. The presence and nature of radio emission at low frequencies is therefore not well-known. For radio emission from the ECMI, low-frequency observations can give insights into electron motions driving the ECMI \citep{2012ApJ...752...60Y}, as well as probing regions of lower magnetic field strength. Using an input magnetospheric structure, low-frequency observations can therefore be used to constrain the coronal volume where conditions are favourable for the ECMI to take place \citep{2018ApJ...854....7L,2011AJ....142..189J}. For gyrosynchrotron emission, the high-frequency regime is typically characterised by a power-law SED, where it is difficult but not impossible to uniquely constrain coronal magnetic field strength, electron density, and other coronal parameters \citep{2017MNRAS.465.1995M, 2016MNRAS.457.1224L, 2015ApJ...802..106L,2011ApJ...735L...2R,2005ApJ...626..486B}. However, at low frequencies the SED transitions to the optically-thick regime, which, when coupled with high-frequency measurements, allows coronal properties to be determined.

Previous low-frequency studies of UCDs include that by \citet{2011AJ....142..189J}, who observed two UCDs with the VLA at 325~MHz, and placed $2.5\sigma$ upper limits around $0.8~\text{mJy}$, and \citet{2016MNRAS.463.2202B}, who observed three nearby UCDs with LOFAR at around 140~MHz, placing $3\sigma$ upper limits around $ 0.7~\text{mJy}$. Unfortunately these limits are not deep enough to provide meaningful constraints on the SEDs of the observed UCDs. Other low-frequency studies of UCDs at low frequencies include a non-detection of TVLM~513-46 at 608 and 1388~MHz by \citet{antonova2007}; and a survey of eight UCDs at 618 and 1288~MHz by \citet{george2009}. Both of these used the Giant Metrewave Radio Telescope (GMRT), and the results were not published in the peer-reviewed literature. 
In this paper, we present new observations of two UCDs observed at $608$ and $1388~\text{MHz}$ with the GMRT taken in 2016. We also present a re-analysis of the 618 and 1288~MHz GMRT observations of eight UCDs first presented in \citet{george2009}. For simplicity, where appropriate we will collectively refer to the set of observations taken at 1288 and 1388~MHz as the 1300~MHz observations, and similarly we will refer to the observations taken at 618 and 608~MHz as the 610~MHz observations.





\section{Observations}
\subsection{Target Selection and Description}
The UCDs we observed during the 2016 campaign (2MASS~J15010818$+$2250020 and 2MASS~J13142039$+$1320011) were selected on the basis of exhibiting stable radio emission over many years. For the 2008 observations first presented by \citet{george2009}, the selected UCDs were a sample of nearby L-dwarfs, as well as the M8.5 dwarf 2MASS~J15010818$+$2250020, and the T6.5 dwarf 2MASS~J00345157$+$0523050. Basic properties of the target UCDs, including the updated coordinates given the proper motion of each UCD, are given in Table \ref{pos_table}. We give a very brief description of each source below.

\textbf{2MASS~J13142039$+$1320011} (also known as NLTT~33370 and LSPM~J1314$+$1320; hereafter J1314$+$1320): this is a binary system consisting of a blended spectral type M7.0e \citep{2006MNRAS.368.1917L}. Its distance of $17.249\pm0.013$~pc determined by \citet{2016ApJ...827...22F} using very long baseline interferometry. It is the most radio-luminous UCD, and has been the subject of many detailed studies at radio frequencies, for example by \citet{2011ApJ...741...27M} and \citet{2015ApJ...799..192W}.

\textbf{2MASS~J15010818$+$2250020} (otherwise known as TVLM~0513$-$46546; hereafter TVLM~513$-$46): this is an M8.5 dwarf at a distance of $10.762\pm 0.027$~pc \citep{2013ApJ...777...70F}. Like J1314$+$1320, it has been the subject of many studies at radio frequencies (e.g. \citealp{2006ApJ...637..518O,2015ApJ...802..106L,2015ApJ...815...64W}), which have revealed both periodic and stochastic flares, as well as quiescent emission observed between $\sim 1$ and 100~GHz.

\textbf{2MASS~J03140344$+$1603056} (hereafter 2M~0314$+$16): this L0.0 dwarf is at a distance of $13.62 \pm 0.05$~pc \citep{2018A&A...616A...1G}. An upper limit on its 8.46~GHz radio luminosity was placed by \citet{2012ApJ...746...23M}.

\textbf{2MASS~J07464256$+$2000321} (hereafter 2M~0746$+$20): this is a near-equal mass binary system of spectral type L0.5 at a distance $12.21\pm0.04$~pc \citep{2009AJ....137....1F}. It was first detected at radio frequencies by \citet{2008A&A...487..317A}, and follow-up studies have detected quiescent and periodic pulsating emission \citep{2009ApJ...695..310B,2015ApJ...802..106L}.

\textbf{2MASS~J08283419$-$1309198} (hereafter 2M~0828$-$13): this L2.0 dwarf is at a distance of $11.69 \pm 0.02$~pc \citep{2018A&A...616A...1G}. Upper limits on its radio luminosity have been placed by \citet{2013A&A...549A.131A} and \citet{2012ApJ...746...23M} at 4.9 and 8.4~GHz respectively.

\textbf{2MASS~J00361617$+$1821104} (hereafter 2M~0036$+$18): this is an L3.5 dwarf at a distance of $8.74\pm 0.01$~pc \citep{2018A&A...616A...1G}. It is a well-studied object at radio frequencies, having been detected, for example, by \citet{2002ApJ...572..503B,2005ApJ...627..960B}, and \citet{2017MNRAS.465.1995M}.

\textbf{2MASS~J03552337$+$1133437} (hereafter 2M~0355$+$11): this L6 dwarf is at a distance of $9.12\pm 0.06$~pc \citep{2018A&A...616A...1G}. An upper limit on its radio luminosity was placed by \citet{2013A&A...549A.131A}. 

\textbf{2MASS~J04234858$-$0414035} (hereafter 2M~0423$-$04): this is a binary system consisting of an L6.5 and a T2 dwarf \citep{2012ApJS..201...19D} at a distance of $14.7\pm0.3$~pc \citep{2018A&A...616A...1G}. Radio pulses from this object were detected by \citet{2016ApJ...818...24K}. 

\textbf{2MASS~J00345157$+$0523050} (hereafter 2M~0034$+$05): this is a T6.5 dwarf at a distance of $9.5 \pm 0.7$~pc \citep{2012ApJ...752...56F}. No radio observations of this object are reported in the literature to date.

\begin{table*}
%
%

%
\centering
\begin{tabular}{ccccccrclc}
\hline
2MASS~J & Short Name & SpT & Epoch & RA (h:m:s) & Dec (d:m:s) & \multicolumn{3}{c}{Distance (pc)} & Reference\\
\hline
\multirow{2}{*}{13142039$+$1320011} & \multirow{2}{*}{J1314$+$1320} & \multirow{2}{*}{M7.0e} & 2016-05-19 & \multirow{2}{*}{13:14:20.1} & \multirow{2}{*}{$+$13:19:57.9} &  \multirow{2}{*}{$17.249$}&\multirow{2}{*}{\hspace{-0.35cm}$\pm$}&\multirow{2}{*}{\hspace{-0.35cm}$0.013$} & \multirow{2}{*}{F16}\\
& & & 2016-05-01 \\
\hline
\multirow{3}{*}{15010818$+$2250020} & \multirow{3}{*}{TVLM~513$-$46} & \multirow{3}{*}{M8.5} & 2016-09-25 & {15:01:08.1} & {$+$22:50:01.1} & \multirow{3}{*}{$10.762$} & \hspace{-0.35cm}\multirow{3}{*}{$\pm$} & \hspace{-0.35cm}\multirow{3}{*}{$0.027$ } & \multirow{3}{*}{F13}\\
& & & 2016-08-18 & {15:01:08.1} & {$+$22:50:01.1} &\\
& & & 2008-01-19 & 15:01:08.2 & $+$22:50:01.7 & \\
\hline
\multirow{2}{*}{03140344$+$1603056} & \multirow{2}{*}{2M~0314$+$16} & \multirow{2}{*}{L0.0} & 2008-01-26& \multirow{2}{*}{03:14:03.3} & \multirow{2}{*}{$+$16:03:05.1} & \multirow{2}{*}{$13.62$} & \multirow{2}{*}{\hspace{-0.35cm}$ \pm$} & \multirow{2}{*}{\hspace{-0.35cm}$ 0.05$} & \multirow{2}{*}{GDR2}\\
& & & 2008-01-25  & \\
\hline
\multirow{2}{*}{07464256$+$2000321} & \multirow{2}{*}{2M~0746$+$20} & \multirow{2}{*}{L0.5} & 2008-01-21& \multirow{2}{*}{07:46:42.3} & \multirow{2}{*}{$+$20:00:31.5} & \multirow{2}{*}{$12.21$} & \multirow{2}{*}{\hspace{-0.35cm}$\pm$} & \multirow{2}{*}{\hspace{-0.35cm}$0.04$} & \multirow{2}{*}{F09}\\
& & & 2008-01-19 &\\
\hline
\multirow{2}{*}{08283419$-$1309198} & \multirow{2}{*}{2M~0828$-$13} & \multirow{2}{*}{L2.0} & 2008-01-21 & \multirow{2}{*}{08:28:33.9} & \multirow{2}{*}{$-$13:09:19.6} & \multirow{2}{*}{$11.69$} & \multirow{2}{*}{\hspace{-0.35cm}$\pm$} & \multirow{2}{*}{\hspace{-0.35cm}$ 0.02$} & \multirow{2}{*}{GDR2} \\
& & & 2008-01-19 & \\
\hline
\multirow{2}{*}{00361617$+$1821104} & \multirow{2}{*}{2M~0036$+$18} & \multirow{2}{*}{L3.5} & 2008-01-28 & \multirow{2}{*}{00:36:16.6} & \multirow{2}{*}{$+$18:21:11.4} & \multirow{2}{*}{$8.74$} & \multirow{2}{*}{\hspace{-0.35cm}$\pm$} & \multirow{2}{*}{\hspace{-0.35cm} $ 0.01$} & \multirow{2}{*}{GDR2}\\
& & & 2008-01-22 & \\
\hline
\multirow{2}{*}{03552337$+$1133437} & \multirow{2}{*}{2M~0355$+$11} & \multirow{2}{*}{L6}   & 2008-01-26 & \multirow{2}{*}{03:55:23.5} & \multirow{2}{*}{$+$11:33:38.6} & \multirow{2}{*}{$9.12$} & \multirow{2}{*}{\hspace{-0.35cm}$\pm$} & \multirow{2}{*}{\hspace{-0.35cm}$ 0.06$}  & \multirow{2}{*}{GDR2}\\
& & & 2008-01-25 & \\
\hline
\multirow{2}{*}{04234858$-$0414035} & \multirow{2}{*}{2M~0423$-$04} & \multirow{2}{*}{L7.5} & 2008-01-24 & \multirow{2}{*}{04:23:48.6} & \multirow{2}{*}{$-$04:14:03.5} & \multirow{2}{*}{$14.7$} & \multirow{2}{*}{\hspace{-0.35cm} $\pm$} & \multirow{2}{*}{\hspace{-0.35cm}$0.3$}  & \multirow{2}{*}{GDR2}\\
& & & 2008-01-18 & \\
\hline
\multirow{2}{*}{00345157$+$0523050} & \multirow{2}{*}{2M~0034$+$05} & \multirow{2}{*}{T6.5} & 2008-01-28 & \multirow{2}{*}{00:34:51.9} & \multirow{2}{*}{$+$05:23:06.4} & \multirow{2}{*}{$9.5 $} & \multirow{2}{*}{\hspace{-0.35cm}$\pm$} & \multirow{2}{*}{\hspace{-0.35cm}$ 0.7$} & \multirow{2}{*}{F12}\\
& & & 2008-01-22 &\\
\hline
\end{tabular}
\caption{\label{pos_table}
Properties of the UCDs analysed in this work. From left to right, the columns indicate: the UCD 2MASS name; the shortened name adopted for this work; spectral type (SpT); the observation epoch UT date; J2000 right ascension and declination at each epoch; distance; and the reference used for the distance, and to calculate updated coordinates. These references are: F16: \citet{2016ApJ...827...22F}; F13: \citet{2013ApJ...777...70F}; GDR2: \citet{2018A&A...616A...1G}; F12: \citet{2012ApJ...752...56F}} 
\end{table*}
\begin{table*}
\begin{tabular}{| c c c c c c |}
\hline
Name & Observation start (UT) &  $\nu$~(MHz) & $\tau$~(h) & Primary calibrator & Secondary calibrator\\
\hline
\multirow{2}{*}{J1314$+$1320} & 2016-05-19 13:00:18 & 1388 & 4.85 & \multirow{2}{*}{3C~286} & \multirow{2}{*}{J1309$+$1104}\\
 & 2016-05-01 15:12:31 & 608 & 5.15 \\
\hline
\multirow{3}{*}{TVLM~513$-$46} & 2016-09-25 12:05:00 & 1388 & 3.97 & 3C~286 & \multirow{3}{*}{1513$+$236}\\
 & 2016-08-18 12:50:03 & 608 & 2.60 & 3C~468 & \\
 & 2008-01-19 12:00:00 & 1288 & 5.43 & 3C~147 & \\
\hline
\multirow{2}{*}{2M~0314$+$16} & 2008-01-26 14:48:04 & 1288 & 1.05 &  \multirow{2}{*}{3C~147} & \multirow{2}{*}{0321$+$123}\\
 & 2008-01-25 11:13:00 & 618 & 2.55 && \\
\hline
\multirow{2}{*}{2M~0746$+$20} & 2008-01-21 14:04:33 & 618 & 3.03 &  \multirow{2}{*}{3C~147} &  \multirow{2}{*}{0738$+$177}\\
 & 2008-01-19 14:10:23 & 1288 & 2.54 & & \\
\hline
\multirow{2}{*}{2M~0828$-$13}  & 2008-01-21 18:29:04 & 618 & 2.65 &  \multirow{2}{*}{3C~147} & \multirow{2}{*}{0902$-$142}\\
& 2008-01-19 17:56:02 & 1288 & 2.62 & \\
\hline
\multirow{2}{*}{2M~0036$+$18} & 2008-01-28 09:15:45 & 618 & 1.67 & \multirow{2}{*}{3C~48} & \multirow{2}{*}{0029$+$349}\\
 & 2008-01-22 07:33:27 & 1288 & 2.52\\
\hline
\multirow{2}{*}{2M~0355$+$11} & 2008-01-26 10:18:47 & 1288 & 1.99 &  \multirow{2}{*}{3C~48} &  \multirow{2}{*}{0321$+$123}\\
& 2008-01-25 11:21:23 & 618 &  2.55 & \\
\hline
\multirow{2}{*}{2M~0423$-$04} & 2008-01-24 13:19:50 & 618 & 3.52 &  \multirow{2}{*}{3C~147} &  \multirow{2}{*}{0423$-$013}\\
& 2008-01-18 13:56:33 & 1288 & 1.87 &  \\
\hline
\multirow{2}{*}{2M~0034$+$05} & 2008-01-28 12:18:37 & 618 & 3.07 &  \multirow{2}{*}{3C~48} & \multirow{2}{*}{0022$+$002}\\ 
 & 2008-01-22 11:29:26 & 1288 & 2.54 & \\
\hline
\end{tabular}
\caption{\label{observation_table}Summary of observations presented in this work. Left to right, the columns indicate: the shortened name (as in Table \ref{pos_table}); the observation start time; central observing frequency $\nu$; total time spent on-source $\tau$; the primary calibrator, and secondary calibrator used during observations.}
\end{table*}
\subsection{Observing Strategy}
We observed J1314$+$1320 and TVLM~513$-$46 with the GMRT in a campaign spanning from May -- September 2016. We conducted observations at two central observing frequencies -- $1388~\text{MHz}$ and $608~\text{MHz}$ -- each with an instantaneous bandwidth of $33.3~\text{MHz}$. These observations were taken using the GMRT Software Backend (GSB).
We observed the primary calibrators in scans lasting approximately 15 minutes at the beginning and end of each observation. Scans centred on the target object lasted approximately 30 minutes and were interleaved between 5-minute duration scans on the secondary calibrator. Primary instrumental polarization products were recorded. Note that the GMRT feeds in the L-band ($\nu = 1-1.45~\mathrm{GHz}$) receiver are linear, and all feeds in lower frequency receivers are circular. In practice, this means that XX and YY instrumental polarizations were recorded for 1300~MHz, where as the 618~MHz observations have RR and LL instrumental polarizations. 

In addition, we also re-analysed observations of eight UCDs presented in the PhD thesis of \citet{george2009}. These observations were carried out in January 2008 using the GMRT Hardware Backend (GHB). Similarly to our 2016 observations, each UCD was observed at $1288~\text{MHz}$ and $618~\text{MHz}$. 
Polarization products were recorded in GHB USB-Polar mode, where RR and LL (618~MHz; circular feeds) or XX and YY (1288~MHz; linear feeds) correlations were both recorded only in the upper $16~\text{MHz}$ of the available bandwidth, and stored in separate \texttt{lta} files. The observing strategy for these observations was very similar our observing strategy for the 2016 observations.

The details of all observations analysed in this work, including the observing frequency, the total time spent on source, and the calibrator sources used, are listed in Table \ref{observation_table}.

\section{Data Reduction}
\label{data_reduction}
We calibrated and imaged the data using the Source Peeling and Atmospheric Modeling (SPAM) Pipeline \citep{ 2014ascl.soft08006I, 2009A&A...501.1185I}. A detailed description of the SPAM pipeline is given by \citet{2017A&A...598A..78I}, but we include a brief description of the pipeline here for completeness. In the first stage of the SPAM pipeline (the `pre-calibration' stage), we performed basic iterative RFI excision, bandpass and complex gain calibration, and flux density scaling. Following this, we performed several rounds of phase-only self-calibration, using an input sky model as an approximate flux density and astrometric reference. We imaged the target field using several facets to account for wide-field imaging effects, and applied several RFI excision routines to the data between rounds of self-calibration and imaging. Following self-calibration, we performed source peeling, and in case of improved signal-to-noise, we saved the self-calibration solutions around the peeled sources. Following the peeling procedure for the brightest field sources, we used the derived solutions to model the time-dependent ionospheric phase delay across the field of view. We subtracted the modelled phase delays from the visibilities, before performing additional rounds of self-calibration to correct any residual phase variations.

We applied the SPAM pipeline to all observations to produce Stokes I continuum images, as well as images for the primary instrumental polarizations. 
In the case of the archival observations, each instrumental polarization was stored in separate \texttt{lta} files. We processed each instrumental polarization separately with SPAM, but supplied a sky model when processing the LL (YY) polarization, by running the \textsc{aegean} source finder \citep{2018PASA...35...11H, 2012MNRAS.422.1812H} on the RR (XX) primary beam-corrected image produced by SPAM at 618~MHz (1288~MHz). This step is justified given that the majority of compact radio continuum sources are unpolarized, meaning that flux densities of most sources from the LL/XX image should be consistent with the RR/YY image. We found that performing this step improved the quality of phase calibrations in some LL data, and at worst did not cause noticeable degradation of the quality of the calibration solutions. Similar phase calibration issues were not identified in the RR data, which is why we chose to calibrate the LL data using the sky model created with RR images. Stokes I continuum images were produced by simply co-adding the images produced for each of the primary instrumental polarizations.

In the 1300~MHz observations (both from 2008 and 2016), there were not enough bright sources in the smaller fields of view to successfully perform the peeling routine with SPAM. In this case, we disabled the peeling process with SPAM. We note that this is not likely to have a significant impact on the quality of the overall calibration, because ionospheric effects are not as significant at frequencies around and higher than 1300~MHz.

In the case of the 2016 observations, we produced Stokes I continuum images directly. If a UCD was detected in the 608~MHz Stokes I images, we split the corresponding LL and RR instrumental polarizations into separate files after the SPAM `pre-calibration' stage. UCDs are well known to exhibit highly polarized bursts of radio emission. To determine if any of the detected UCDs were significantly circularly polarized, we processed the RR and LL instrumental polarizations individually with SPAM. We used the method implementing a sky model with SPAM as outlined above, but for the 2016 observations the sky model was produced from the full Stokes I continuum image rather than a single instrumental polarization. 

All observations were imaged with robust weighting. The centrally condensed GMRT array can lead to strong, broad sidelobes around bright sources in the field. We set the briggs parameter to $-1$, being closer to uniform weighting, therefore reducing the impact of sidelobe confusion and large-scale sidelobe artefacts in the final images.

\begin{figure*}	
\includegraphics[width=0.493\textwidth]{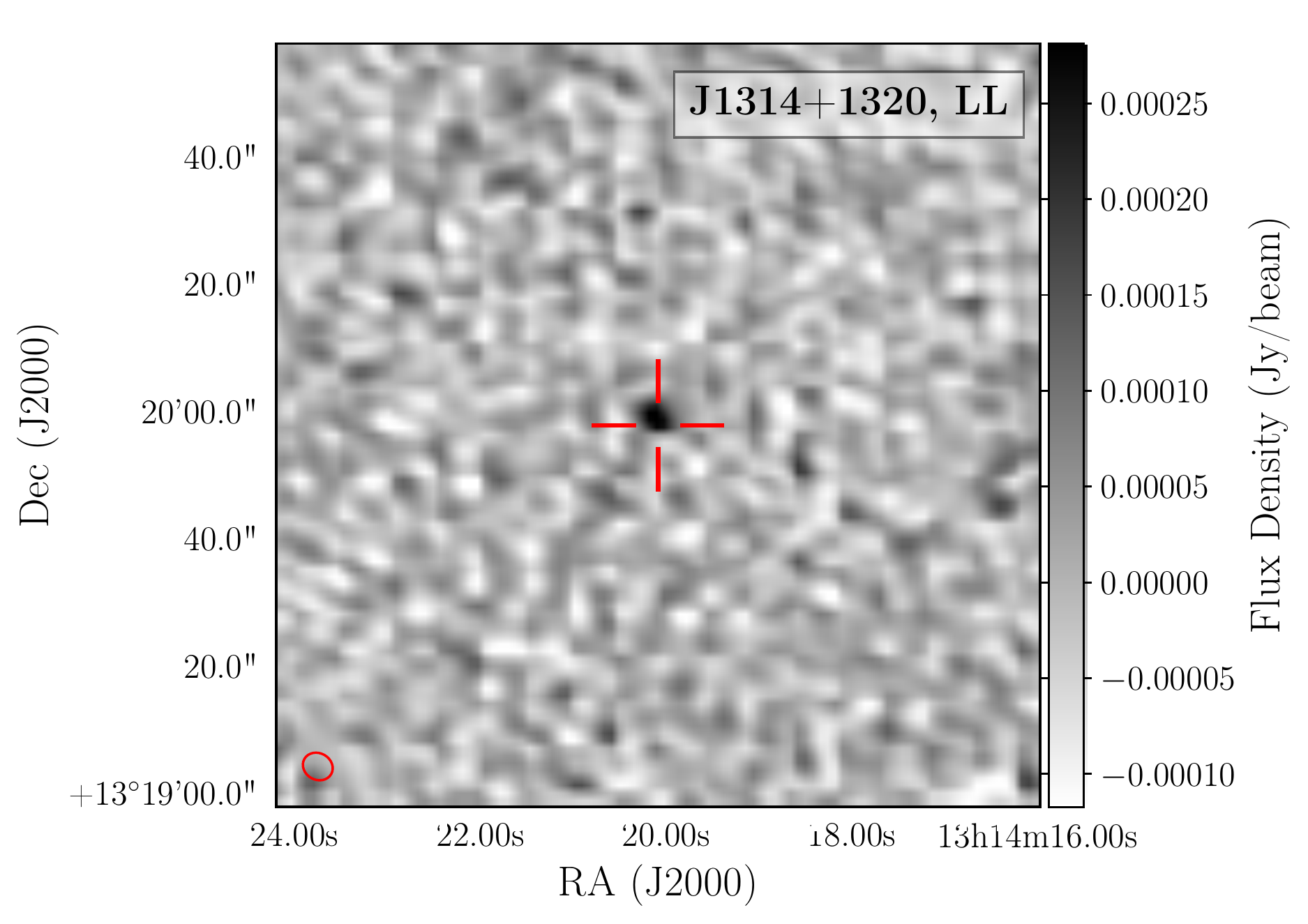}
\includegraphics[width=0.493\textwidth]{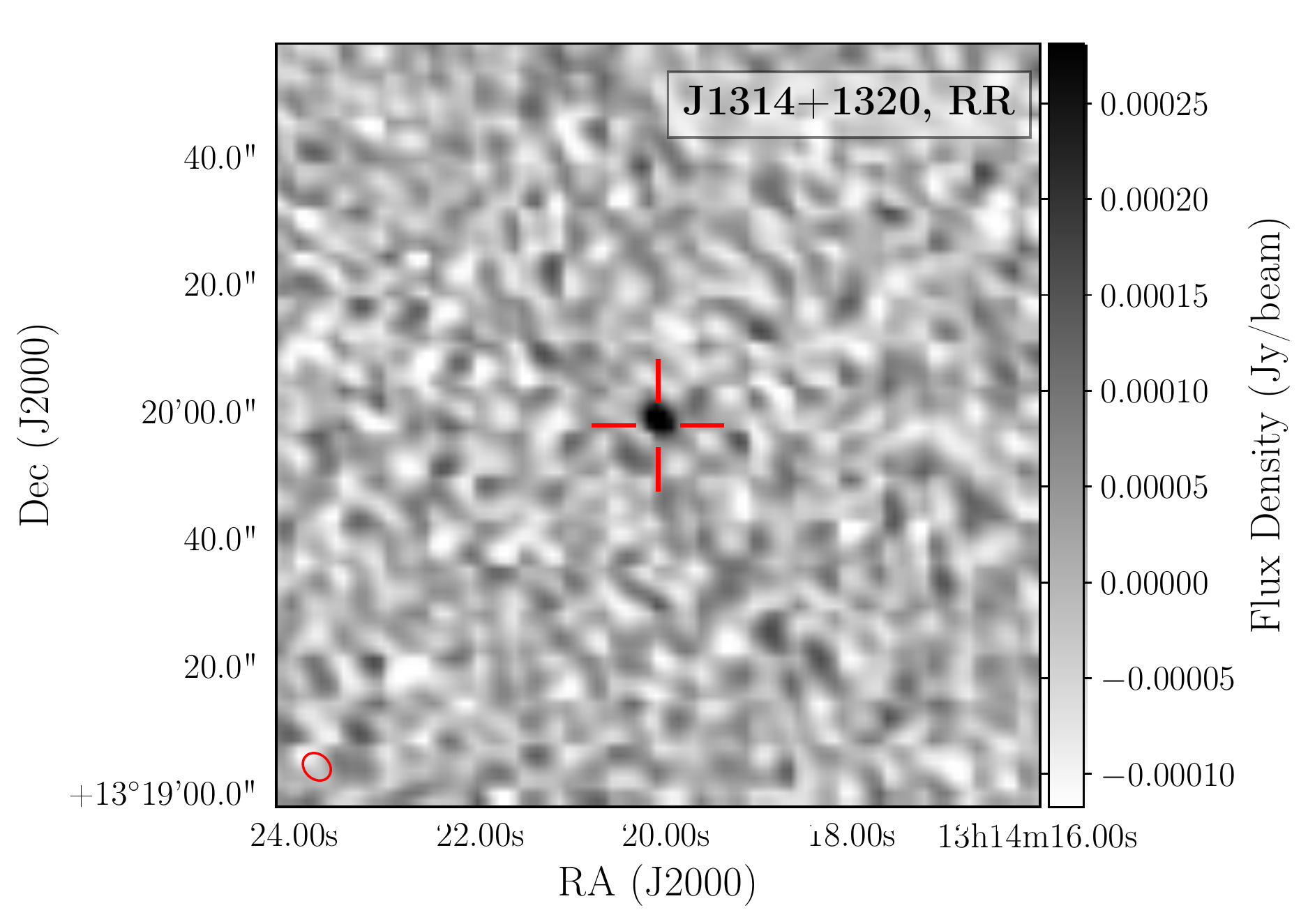}
\caption{\label{J1314+1320_LL_RR}Postage stamp images of J1314$+$1320, in LL (left sub-figure) and RR (right sub-figure) instrumental polarizations, at an observing frequncy of 608~MHz. The expected location of the UCD, given its proper motion, is indicated by the red crosshairs. The shape of the synthesised beam is indicated by the red ellipse in the lower left corner. A detection of a point source at the expected location of J1314$+$1320 is clear in both the LL and RR images.}
\end{figure*}

\begin{figure*}
\includegraphics[width=0.493\textwidth]{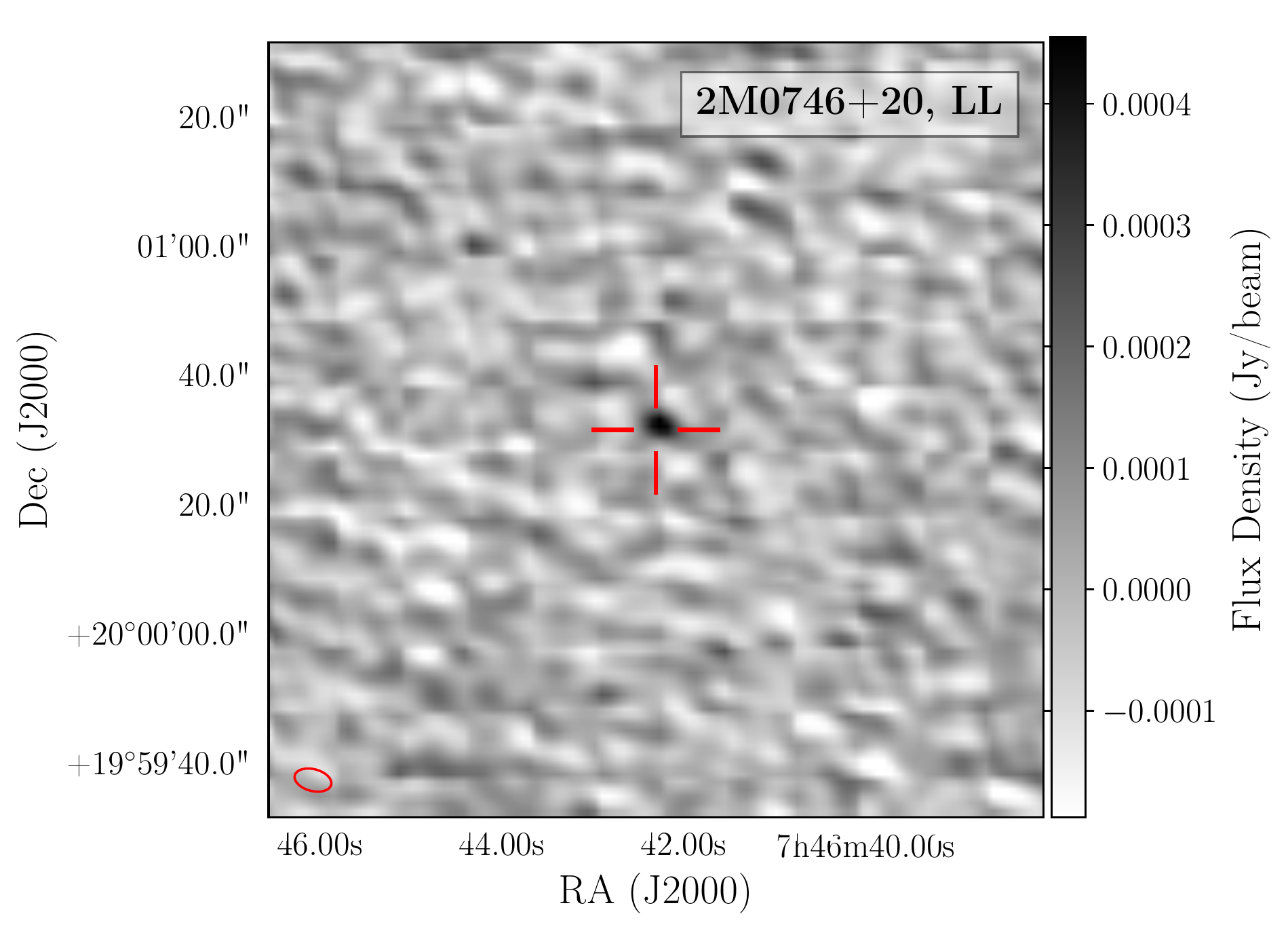}
\includegraphics[width=0.493\textwidth]{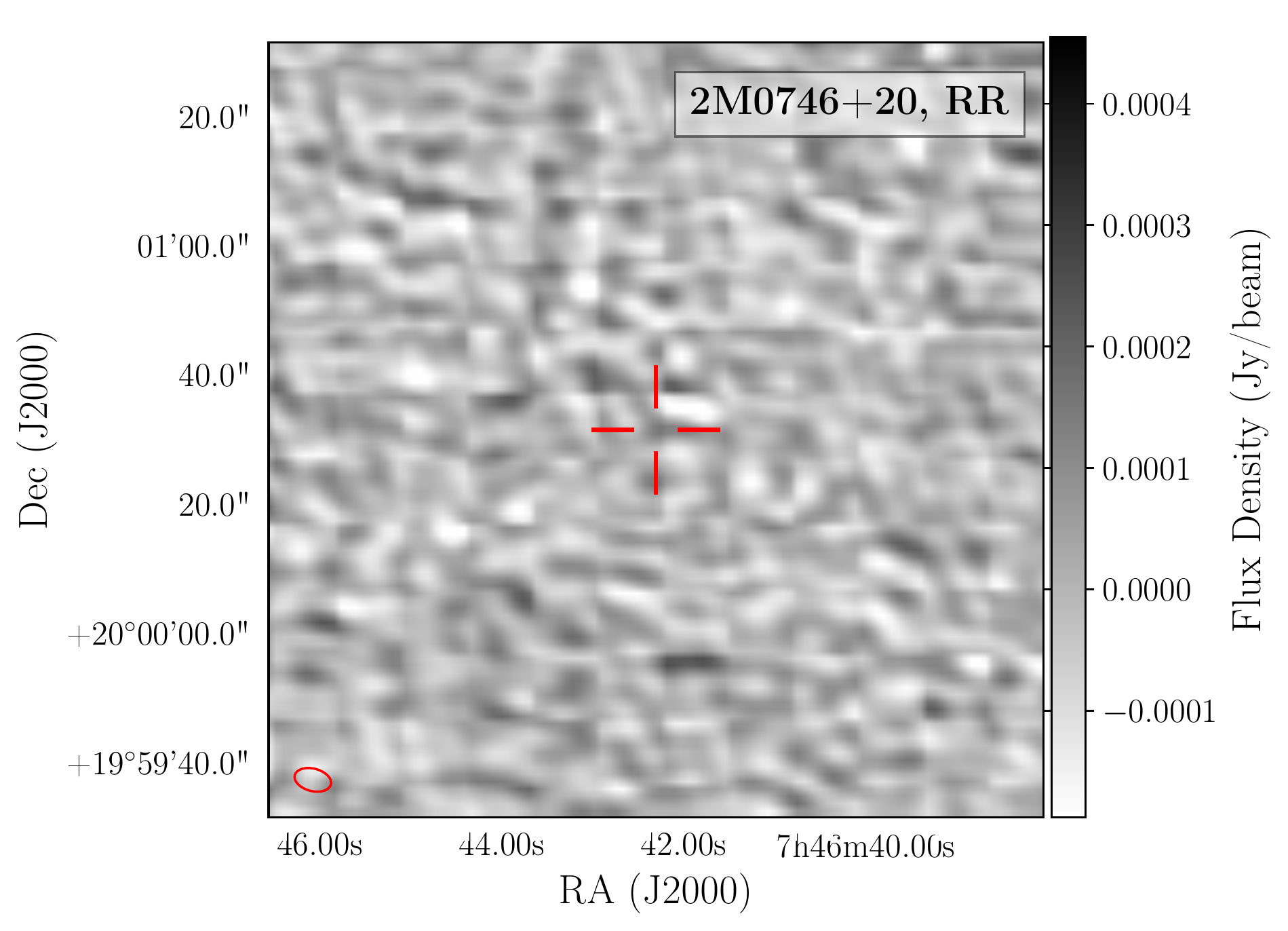}
\caption{\label{2M0746+20_LL_RR} Same as Figure \ref{J1314+1320_LL_RR}, but for 2M~0746$+$20 at 618~MHz. The high degree of left-circular polarization is revealed by the strong detection ($481 \pm 69~\mu\mathrm{Jy}$) in the LL instrumental polarization, and non-detection in RR, with a $3\sigma$ upper limit of $241~\mu\mathrm{Jy}$.}
\end{figure*}

\section{Results}
\begin{table}
\begin{tabular}{clllll}
\hline\\
Name & $S_{610}$ & $S_{1300}$ & $T_{b,610}$ & $T_{b,1300}$ \\
 & ( $\mathrm{\mu Jy}$ ) & ( $\mathrm{\mu Jy}$ ) & ( $10^{10} \mathrm{K}$ ) & ( $10^{10} \mathrm{K}$ )\\
 \hline\\
 J1314$+$1320 & $391 \pm 26$  & $908 \pm 65$ & $ 19 \pm 1 $       & $ 8.8 \pm 0.6 $ \\
 \multirow{2}{*}{TVLM~513$-$46} & \multirow{2}{*}{$<148$} $^{\mathrm{a}}$ & {$<194$} $^{\mathrm{b}}$ & \multirow{2}{*}{$<2.8$} $^{\mathrm{a}}$ & {$<0.85$} $^{\mathrm{b}}$\\
                             &                                        & {$<201$} $^{\mathrm{c}}$&                          & {$<0.76$} $^{\mathrm{c}}$\\
2M~0314$+$16  & $<200$       & $<728$       & $<6.3 $            & $<5.1 $  \\
2M~0746$+$20  & $319 \pm 58$ & $355 \pm 65$ & $ 8.0 \pm 1.5 $       & $ 2.0 \pm 0.4 $ \\
2M~0828$-$13  & $<160$       & $<165$       & $<3.7 $            & $<0.85  $  \\
2M~0036$+$18  & $<800$       & $<1040$      & $<10 $             & $<3.0 $ \\
2M~0355$+$11  & $<184$       & $<1020$      & $<2.6 $            & $<3.2 $  \\
2M~0423$-$04  & $<150$       & $<1360$      & $<5.5 $            & $<11  $ \\
2M~0034$+$05  & $<245$       & $<243$       & $<3.8 $            & $<0.83 $ \\
\hline\\
\end{tabular}
\caption{\label{quiescent_results}
Summary of detections and $3\sigma$ upper limits of the observed UCDs. Left to right, the columns indicate the shortened UCD name; the flux density at 610~MHz; flux density at 1300~MHz; and the corresponding brightness temperatures at 610 and 1300~MHz.
Notes: (a): Epoch 2016-08-18; (b): Epoch 2008-01-19; (c): Epoch 2016-09-25}
\end{table}
\subsection{Non-flaring emission}
In each of the continuum images, we used \textsc{casa} task \texttt{imfit} to determine the integrated flux density at the expected location of each UCD, given measurements of proper motion reported in the literature. In the case of a detection, we report uncertainties in flux density as computed in the \texttt{imfit} process. Where there was no detection, we calculated the RMS noise in the local region around the UCD using \textsc{casa} task \texttt{imstat}, and used this to set $3\sigma$ upper limits on the flux density. A summary of the results for non-flaring emission for each UCD is given in Table \ref{quiescent_results}. Of the nine UCDs, we detect radio emission from two: J1314$+$1320, and 2M~0746$+$20. These objects are detected both at $\sim 610$ and 1300 MHz. Postage-stamp images showing the detections of J1314$+$1320 and 2M~0746$+$20 at 610~MHz, in LL and RR instrumental polarizations, are shown in Figures \ref{J1314+1320_LL_RR} (J1314$+$1320) and \ref{2M0746+20_LL_RR} (2M~0746$+$20).


\subsection{Variable emission}
To search for any variable or flaring emission, we imported the SPAM-calibrated Stokes I visibilities into \textsc{casa} v5.1.2 \citep{2007ASPC..376..127M} and imaged  the full length of the observations. We subtracted the CLEAN model components for all field sources except for the target UCD from the visibilities using \textsc{casa} task \texttt{uvsub}. We imaged the residual $uv$-subtracted data on 2, 5, and 10-minute time-scales using \textsc{casa} task \texttt{tclean}. Similarly to the continuum images we produced using SPAM, we produced short-time-scale images using robust weighting with a briggs factor of $-1$. We created light curves by taking the peak flux density in a box centred at the expected location of each UCD, with dimensions equal to 2.5 times the semi-major axis of the restoring beam in the image. The dimensions of the fitting box were slightly enlarged to account for any potential spurious astrometric offsets produced during SPAM processing. We calculated the local RMS noise by taking the standard deviation in a large box around the expected location of the source, clipping out pixels with flux densities deviating by more than $3\sigma$, and evaluating the standard deviation of the residual pixels. We set a detection threshold of $5\sigma$ for any flaring or variable emission. We did not find any emission above this threshold over 2, 5, and 10-minute time-scales, indicating a lack of flaring activity of these UCDs at these frequencies, at levels above a few millijanskys. We did not fold the light-curves at the expected rotational period of each UCD. This is because our observations are only long enough to cover approximately two periods at best, meaning that phase-folding only provides a modest gain in sensitivity.

\subsection{Characterising the radio emission}
\subsubsection{Brightness temperatures}
\label{brightness_temperatures}
To further characterise the observed radio emission, we calculated the corresponding brightness temperatures using the expression \begin{equation}
T_b = 2.5 \times 10^{9} \left(\frac{S_\nu}{\mathrm{mJy}}\right) \left(\frac{\nu}{\mathrm{GHz}}\right)^{-2}\left( \frac{d}{\mathrm{pc}}\right)^{2}\left( \frac{R}{R_\mathrm{Jup}} \right)^{-2}~\mathrm{K},
\end{equation}
with flux density $S_\nu$ at frequency $\nu$, measured from a source of distance $d$, and emission region of radius $R$. Here, $R_\mathrm{Jup} = 7.0\times 10^{9}~\mathrm{cm}$ is the radius of Jupiter, which is thought to be the typical radius of UCDs \citep{2001RvMP...73..719B}. There have been no direct measurements of the radio emission region size of UCDs. We therefore adopted an emission region radius of $2~R_*~( = 2~R_\mathrm{Jup})$, thought to be the size of M-type stellar coronae \citep{2000A&A...359.1035L}. We find that the measured flux densities and upper limits yield brightness temperatures in the range of $(0.76 - 19)\times 10^{10}~\mathrm{K}$. For the detected UCDs, these brightness temperatures rule out thermal bremsstrahlung emission as the source of the radio emission, favouring non-thermal emission processes instead. In a following section (Section \ref{coronal_modelling}), we describe modelling of the corona of J1314$+$1320. When performing this modelling, we allowed the total radio emission region size $L$ (where $L = 2R$) to vary between $1.0-12.0~R_*$. Adopting these values alters the brightness temperatures by factors between $16$ and $1/9$ relative to our brightness temperature calculated with $R = 2~R_*$. Regardless, the brightness temperatures of the detected UCDs remain greater than $2.2\times10^9~\mathrm{K}$, inconsistent with thermal emission.

\subsubsection{Polarization}
For the detections made at 610 MHz, we were able to determine the degree of circular polarization by calibrating and imaging the LL and RR instrumental polarizations independently, as described in Section \ref{data_reduction}.  Because SPAM does not perform any polarization calibration, our detections may be subject to leakage on the order of $5$ per cent \citep{2014arXiv1407.0528F}. We used \textsc{casa} task \texttt{imfit} to determine the flux density of the target source in each instrumental polarization.

For J1314+1320, we find that the LL flux density at 608~MHz is $357\pm 44~\mu\mathrm{Jy}$, and the RR flux density is $523\pm 46~\mu\mathrm{Jy}$. Because our target source is close to the phase centre, the corresponding leakage should be relatively small \citep{2014arXiv1407.0528F}. To estimate the polarization leakage, we measured the flux density of two nearby bright sources in the LL and RR images. Using these flux densities, we calculated polarization fractions of $2 \pm 2$ per cent. We add a conservative leakage estimate of $5$ per cent in quadrature with the relevant measurement uncertainties. This implies a circular polarization fraction $\pi_c = (S_\mathrm{RR} - S_\mathrm{LL})/(S_\mathrm{RR} + S_\mathrm{LL})$ of $+19 \pm 12$ per cent. Here, $S_\mathrm{RR}$ and $S_\mathrm{LL}$ are the RR and LL flux densities, respectively.

For 2M~0746+20, we measure a flux density in the 618~MHz LL instrumental polarization of $481 \pm 69~\mu\mathrm{Jy}$, but in RR, we make no detection, with a $3\sigma$ upper limit of $241~\mu\mathrm{Jy}$. We calculate a $3\sigma$ upper limit on the polarization fraction as follows. We take the $3\sigma$ upper limit of $241~\mu\mathrm{Jy}$ as the fiducial measured value of the RR flux density, for the purpose of calculating the appropriate uncertainty only. We then apply standard error propagation using the measured uncertainties and flux densities to calculate an uncertainty in the fiducial polarisation fraction of 24.5 per cent. Multiplying this by -3 (where the minus sign indicates a left-handed sense of circular polarization, following convention) gives a $3\sigma$ upper limit of $\pi_c < -76$ per cent. We note that because there is no detection in the RR instrumental polarization, there is no need to take leakage into account for this lower limit.

\subsubsection{Radio emission mechanism}
As mentioned in Section \ref{brightness_temperatures}, the high brightness temperatures of the detected UCDs rule out thermal emission being responsible for the observed radio emission. However, they do not exceed the $10^{12}~\mathrm{K}$ limit for the inverse-Compton catastrophe \citep{1969ApJ...155L..71K}. Although some (e.g. \citealp{2008ApJ...684..644H}) have argued that steady emission from UCDs may be generated by the ECMI, the inferred brightness temperatures do not favour coherent emission processes. We therefore cannot rule out incoherent emission processes such as gyrosynchrotron radiation from a non-thermal electron population.  


For J1314$+$1320, the modest level of circular polarization (approximately $14$ per cent) also disfavours coherent emission processes which can have polarization fractions up to $100$ per cent. Furthermore, the detection of persistent radio emission across a very broad frequency range, from $608~\mathrm{MHz}$ to  $22.5~\mathrm{GHz}$ \citep{2011ApJ...741...27M} suggests that the ECMI is not the source of the observed emission, since ECMI emission is expected to be strongly peaked around the cyclotron frequency and its harmonics. Finally, the positive spectral index at low frequencies is suggestive of optically-thick gyrosynchrotron radiation, and we conclude that this is likely to be the source of the persistent emission observed from this source.

For 2M~0746+20, the upper limit on the fractional circular polarization of -74 per cent is quite strong, but is still consistent with gyrosynchrotron radiation (e.g. \citealp{1985ARA&A..23..169D,2016ApJ...822...34P}). The perceived lack of variability, along with the broad frequency range over which flux densities have been recorded ($610~\mathrm{MHz}$ to $8.5~\mathrm{GHz}$) do not favour ECMI as the emission source. Therefore, we conclude that these observations mildly favour gyrosynchrotron radiation as the emission mechanism, though we cannot strongly rule out ECMI as being the emission mechanism, primarily due to the strong upper limit of fractional circular polarization $\pi_c < -74$ per cent. Future broad-band, simultaneous radio observations of this star would assist in clarifying the nature of its radio emission mechanism.

\begin{figure}
\includegraphics[width=\linewidth]{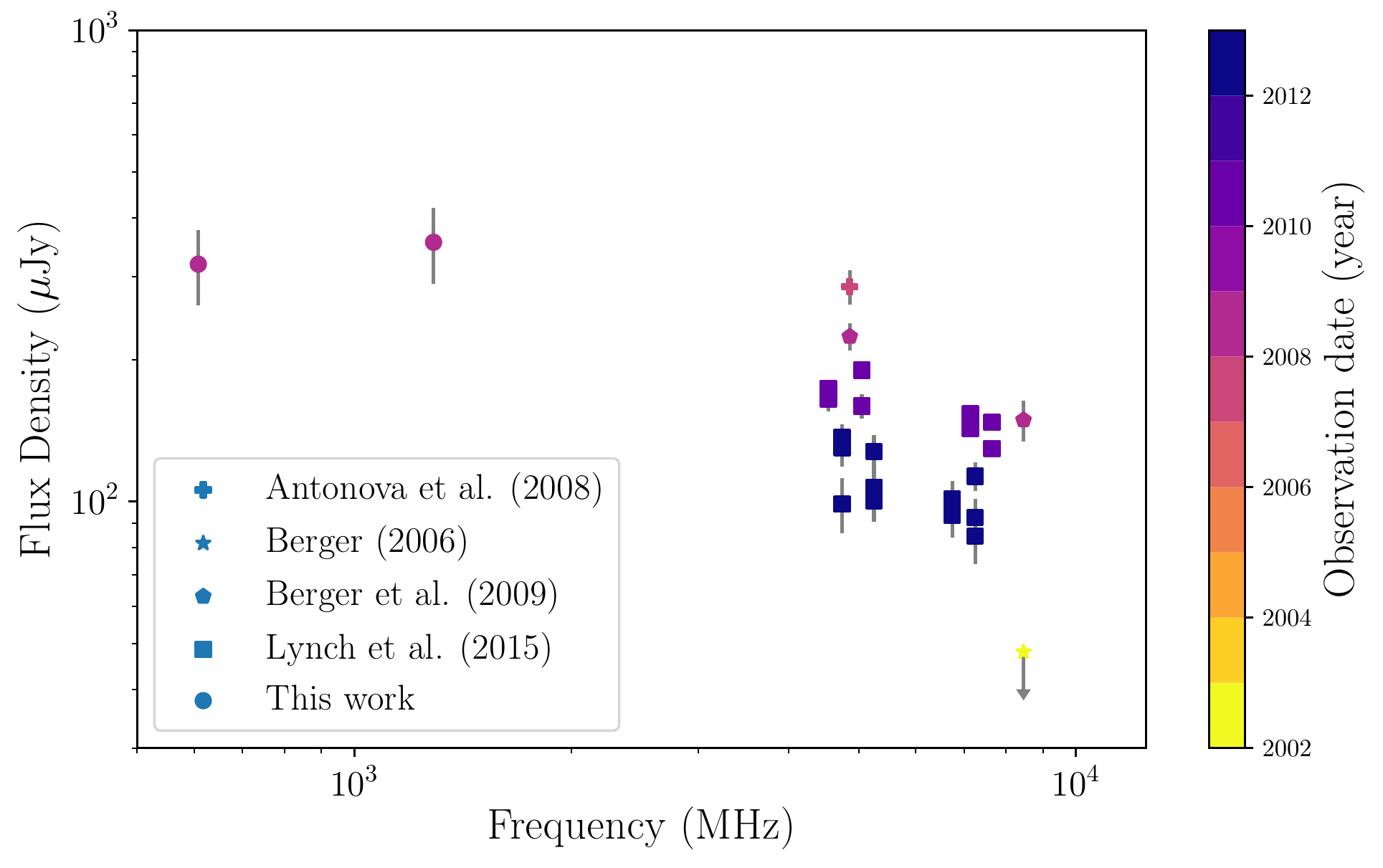}
\caption{\label{2M0746+20_SED}Non-simultaneous SED of 2M~0746$+$20, with flux densities as measured in this work, and taken from \citet{2008A&A...487..317A, 2006ApJ...648..629B,2009ApJ...695..310B}, and \citet{2015ApJ...802..106L}. A high degree of long-term variability at high frequencies, along with an unclear turnover frequency, means that we cannot use the spectral turnover frequency as an additional constraint on the coronal parameters of this UCD. We therefore do not attempt to model the coronal parameters of this UCD.}
\end{figure}
\begin{figure}
\includegraphics[width = \linewidth]{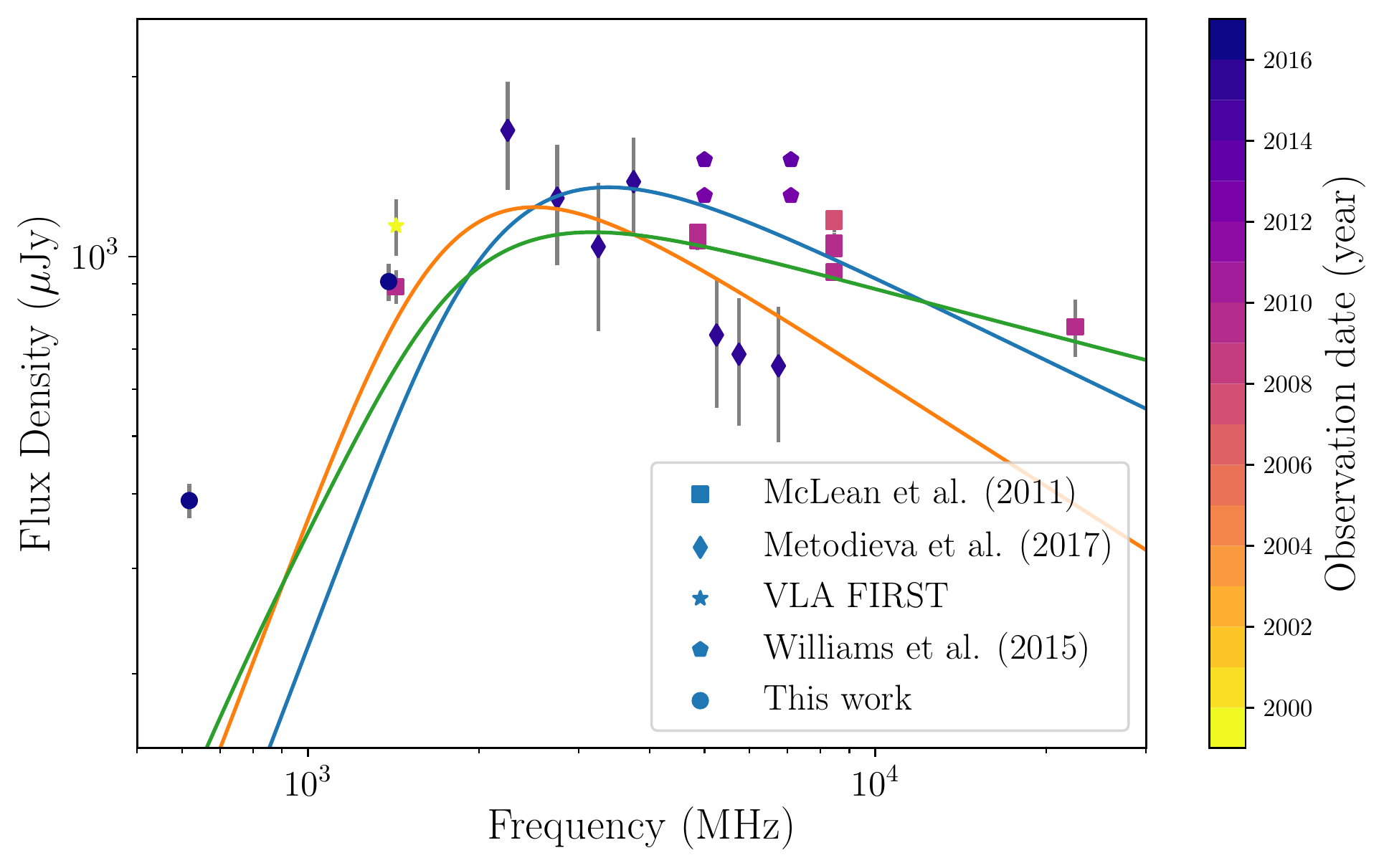}
\caption{\label{J1314_sed}Non-simultaneous SED of J1314$+$1320, with flux densities as measured in this work, and taken from \citet{2011ApJ...741...27M,2015ApJ...799..192W} and \citet{2017MNRAS.465.1995M}. Using these measurements, we produced model gyrosynchrotron SEDs, varying the radio emission region size $L$, electron energy power-law index $\delta$, coronal magnetic field strength $B$, and electron number density $N_e$.
Some example model SEDs for coronal parameters consistent with the observed flux densities are shown as blue, orange, and green solid lines. Respectively, these correspond to coronal parameters of:
$L = 5.0~R_*, \delta = 1.91, B = 1.5~\mathrm{G}, N_e = 2 \times 10^{8}~\mathrm{cm}^{-3}$ (blue);
$L = 8.0~R_*, \delta = 2.19, B = 2.8~\mathrm{G}, N_e = 9 \times 10^{7}~\mathrm{cm}^{-3}$ (orange);
$L = 11.0~R_*, \delta = 1.47, B = 18.0~\mathrm{G}, N_e = 4 \times 10^{4}~\mathrm{cm}^{-3}$ (green). These example model SEDs are qualitatively consistent with the observed spectral flux densities, and illustrate that a wide range of coronal parameters are plausible for J1314$+$1320.
}
\end{figure}

\subsection{Coronal modelling}
\label{coronal_modelling}
\begin{figure}
\includegraphics[width=\linewidth]{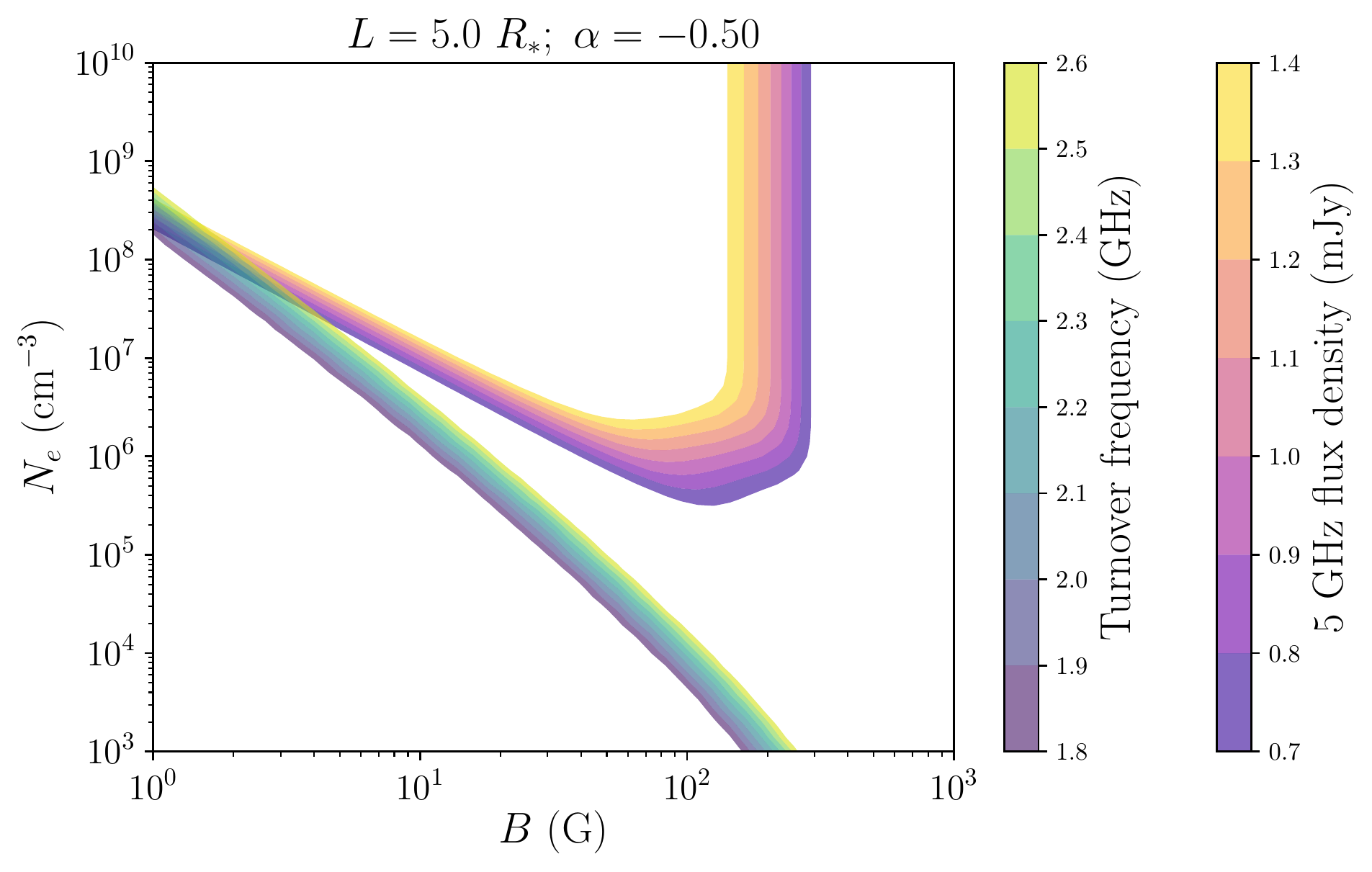}
\includegraphics[width=\linewidth]{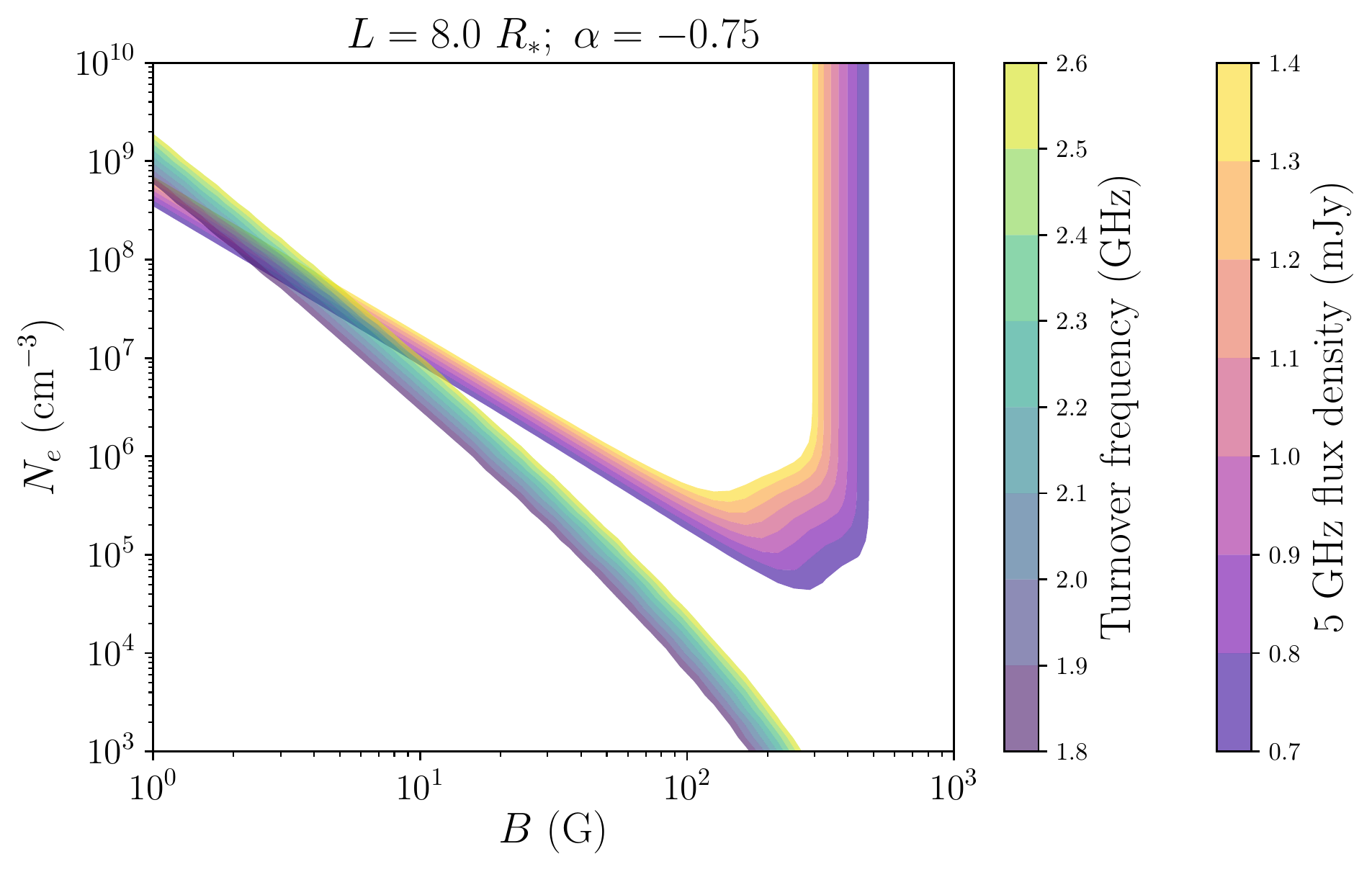}
\includegraphics[width=\linewidth]{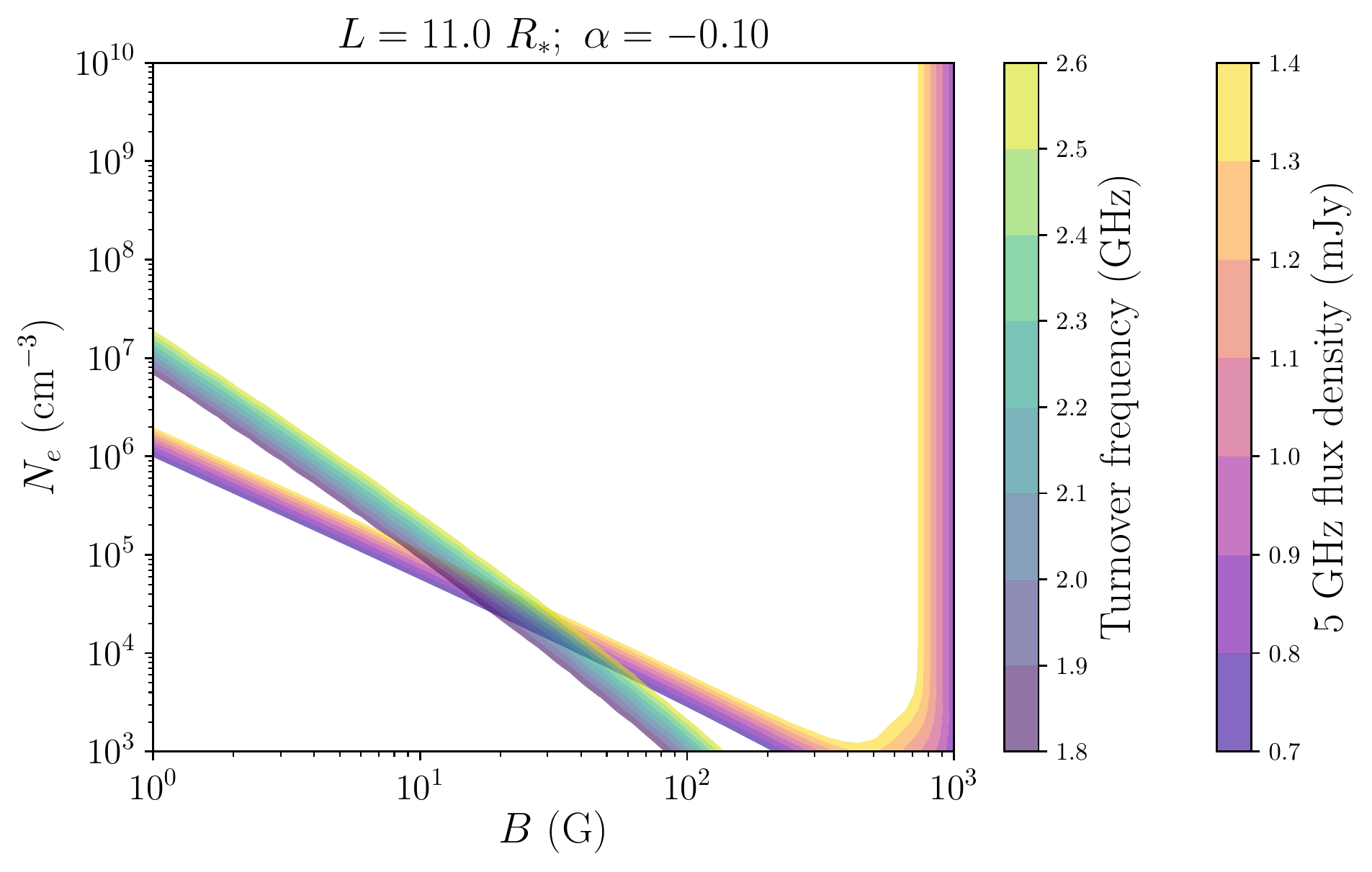}
\caption{\label{J1314+1320_params}Representative plots showing regions of $N_e$ and $B$ parameter space consistent with criteria on the spectral turnover of J1314$+$1320 being close to $2.2~\mathrm{GHz}$, and 5~GHz flux density around $1~\mathrm{mJy}$. The region of parameter space where the two sets of contours overlap is where the parameters are consistent with our observations. The shaded contours represent turnover frequencies and 5~GHz flux densities for
$L = 5.0~R_*,~\alpha_\mathrm{thin} = -0.5$; $L = 8.0~R_*,~\alpha_\mathrm{thin} = -0.75$; $L = 11.0~R_*,~\alpha_\mathrm{thin} = -0.1$ from top to bottom, respectively.}
\end{figure}
\begin{table}
\centering
\def\arraystretch{2}\tabcolsep=2pt
\begin{tabular}{c|lllll}
\hline
Work & $\alpha_\mathrm{thick}$ & $\alpha_\mathrm{thin}$ & $S_\mathrm{br}~\mathrm{(mJy)}$ & $\nu_\mathrm{br}~\mathrm{(GHz)}$ \\
\hline
M17 & $1.04^{+0.12}_{-0.13}$ & $-0.90^{+0.27}_{-0.32}$ & $1.57^{+0.17}_{-0.18}$ & $2.36^{+0.30}_{-0.26}$ \\
W15 & $1.01^{+0.12}_{-0.13}$ & $0.00 \pm 0.02$ & $1.36\pm0.03$ & $2.1^{+0.3}_{-0.2}$ \\
M11 & $1.0\pm0.1$ & $-0.10\pm0.03$ & $1.21\pm0.04$ & $1.93^{+0.16}_{-0.13}$ \\
All data & $1.0 \pm 0.1$ & $-0.17\pm0.02$ & $1.52\pm0.03$ & $2.30 ^{+0.16}_{-0.13}$ \\
\hline
\end{tabular}
\caption{\label{bpl_table}Results from the MCMC fit of a broken power-law to the non-simultaneous flux density measurements across the SED of J1314$+$1320. Each row indicates the results from a fit to the low frequency GMRT measurements paired with the measurements from the works cited in the first column. The final row shows the results from combining all measurements to use in MCMC fitting. From left to right, the columns indicate the work where measurements are taken from; the optically thick (low-frequency) spectral index ($\alpha_\mathrm{thick}$); the optically-thin (high-frequency) spectral index ($\alpha_\mathrm{thin}$); the flux density at the spectral turnover $S_\mathrm{br}$; and the spectral turnover frequency $\nu_\mathrm{br}$. Works used: M17: \citet{2017MNRAS.465.1995M}; W15: \citet{2015ApJ...799..192W}; M11: \citet{2011ApJ...741...27M}}
\end{table}
The flux density measurements across the SED of each UCD are non-simultaneous, with gaps of months or years between individual observations. It is well-known that the non-flaring emission from UCDs can vary significantly over long time-scales (e.g \citealp{2007A&A...472..257A}). Therefore, care should be taken when estimating coronal properties of UCDs using non-simultaneous spectral flux density measurements. 
Figure \ref{2M0746+20_SED} shows that the spectral flux densities of 2M~J0746$+$20 vary by factors of a few over long time-scales. Moreover, while our low-frequency flux density measurements hint at a rising spectrum (increasing flux density with increasing observing frequency), the location of the turnover frequency is not clear. For instance, the difference in flux densities at 618 and 1288~MHz is sufficiently small to fully explained by long-term variability of the quiescent emission. We therefore cannot use the spectral turnover frequency as an additional constraint on this UCD's coronal properties, and so we refrain from modelling its coronal properties in this work.

On the other hand, as shown in Figure \ref{J1314_sed}, long-term variability of the persistent emission from J1314$+$1320 appears to be more mild than that of 2M~0746$+$20. Additionally, the 608~MHz flux density unambiguously shows a rising spectrum at low frequencies. These features allow us to make order-of-magnitude estimates of the coronal properties of J1314$+$1320. None the less, because of the uncertainty introduced by long-term variability, we used a simple coronal model, containing a uniform power-law electron density within a homogeneous magnetic field of strength. We used the expressions for gyrosynchrotron emission and absorption coefficients given in \citet{1984AuJPh..37..675R} for a population of electrons with a power-law energy distribution of the form \begin{equation}
N_e(\gamma) = N_0(\delta - 1)E_0^{\delta - 1}(\gamma - 1)^{\delta}
,\end{equation}
where $\gamma$ is the electron Lorentz factor, $\delta$ is the electron power-law index, $N_0 = \int_1^\infty N(\gamma) d\gamma$ is the total non-thermal electron population, and $E_0$ is the low-energy cutoff ($N(\gamma) = 0$ when $(\gamma - 1) < E_0$). We performed radiative transfer simulations to calculate the corresponding spectral flux density for a given choice of model parameters. We varied the coronal magnetic field strength $B$, electron number density $N_e$, and emission region size $L$. Note that $\delta$ depends on the optically-thin spectral index $\alpha_\mathrm{thin}$ through
\begin{equation}
\delta = -\frac{\alpha_\mathrm{thin} - 1.22}{0.9},
\end{equation} \citep{1985ARA&A..23..169D}. Figure \ref{J1314_sed} shows that the optically-thin spectral index of J1314$+$1320 varies over long timescales. Therefore, we also varied the electron power-law index $\delta$ in the radiative transfer modelling.
We found that the modelled flux densities depend weakly on the viewing angle $\theta$ between the line of sight and the magnetic field, for values between $10-80^\circ$. Therefore, to simplify the modelling procedure, we chose to keep $\theta$ fixed at $25^{\circ}$, within the range of acceptable values reported by \citet{2017MNRAS.465.1995M}, ignoring the sense of polarization.

We empirically determined the spectral turnover frequency by running a Markov Chain Monte Carlo sampler, from the \texttt{emcee} package \citep{2013PASP..125..306F}, to fit a broken power law of the form 
\begin{equation}
S_\nu = \left\{
        \begin{array}{ll}
            S_\mathrm{br}(\nu/\nu_\mathrm{br})^{\alpha_\mathrm{thick}} & \quad \mathrm{if }~\nu \leq \nu_\mathrm{br} \\
            S_\mathrm{br}(\nu/\nu_\mathrm{br})^{\alpha_\mathrm{thin}} & \quad \mathrm{if }~\nu > \nu_\mathrm{br}
        \end{array}
        \right. ,
\end{equation}
where $\alpha_\mathrm{thick}$ is the optically-thick spectral index, $S_\mathrm{br}$ is the flux density at the spectral break frequency $\nu_\mathrm{br}$, representing the transition frequency from optically thick to optically thin emission. We used a uniform prior on the break frequency, setting bounds of $1.0 \leq \nu_\mathrm{br} \leq 4.86$~GHz.

We fit the broken power law to the flux density measurements reported in this work, and from \citet{2017MNRAS.465.1995M}, \citet{2015ApJ...799..192W}, and \citet{2011ApJ...741...27M}. Because the measurements reported in \citet{2017MNRAS.465.1995M} and \citet{2015ApJ...799..192W} were taken simultaneously, we paired our low-frequency measurements with the measurements from these two works, as well as from \citet{2011ApJ...741...27M}, separately. We fit the data independently for each pairing, as well as producing an overall fit using all measurements. This enabled us to account for long-term spectral variability in our analysis.
The best-fitting parameters using measurements from each work are given in Table \ref{bpl_table}. 

To qualitatively assess the goodness of fit of a model to the measured flux densities, we allow models that give turnover frequencies between 1.8 and 2.6 GHz, corresponding with the extreme upper and lower bounds of all measurements of $\nu_\mathrm{br}$. The modelled turnover frequencies were determined by finding where the optical depth was equal to 1. For comparison of our modelled flux densities to the measurements at around $5~\mathrm{GHz}$, we allow models producing $5~\mathrm{GHz}$ flux densities between 0.7 and 1.4~mJy. We chose not to compare the model to our low-frequency measurements because the numerical expressions for emission and absorption coefficients given in \citet{1984AuJPh..37..675R} break down when the ratio of the observing frequency to the plasma cyclotron frequency is of order unity or less. This occurs for observing frequencies of a few hundred megahertz, with magnetic field strengths of a few hundred gauss.

Some representative plots showing regions of parameter space agreeing with our criteria on the turnover frequency and 5~GHz flux density are shown in Figure \ref{J1314+1320_params}. For a fixed value of emission region size $L$ and electron power-law index $\delta$ we are able to constrain $N_e$ and $B$ to within one or two orders of magnitude. However, these constraints on $N_e$ and $B$ are sensitive to the choice of $L$ and $\delta$, so overall we are only able to constrain these parameters to a wide range. For an emission size $3.0 < L < 12.0~R_*$ and $1.36 < \delta < 2.47$ (corresponding to $-1.0 < \alpha_\mathrm{thin} < 0.0$), we are able to constrain $1 \lesssim B \lesssim 90~\mathrm{G}$ and $4 \lesssim \log(N_e) \lesssim 10$. These results are consistent with those found by \citet{2017MNRAS.465.1995M}. We find that for $L < 3~R_*$, $B \ll 1~\mathrm{G}$ and $\log(N_e) \gg 8$, which we rule out as physically unrealistic parameter combinations. Our parameter estimates are clearly degenerate -- general trends we find between the parameters are that larger emission region sizes and shallower spectral indices tend to correspond with higher magnetic field strengths and lower electron number density. Some example model SEDs for coronal parameters consistent with the observed spectral flux densities are shown overlaid with the flux density measurements in Figure \ref{J1314_sed}. Our results none the less re-illustrate (as in \citealt{2017MNRAS.465.1995M}) that precise coronal parameters can be determined with simultaneous low and high-frequency radio measurements, and if the source emission size can be determined independently.

\section{Summary and Future Work}

We have analysed observations of nine UCDs at $\sim 610$ and $1300~\mathrm{MHz}$ taken with the GMRT, in order to characterise the low-frequency radio emission of UCDs. We find no evidence for flaring activity at a level above a few millijanskys. However, we observe persistent radio emission from J1314$+$1320 and 2M~0746$+$20 at $\sim$610 and 1300~MHz, making these the lowest-frequency detections of these sources to date. We find no evidence for significant variability of the persistent emission.

The high brightness temperature and moderate polarization indicate that gyrosynchrotron radiation from a power-law distribution of mildly-relativistic electrons is the most likely source of the radio emission observed from these UCDs. We apply radiative transfer modelling to J1314$+$1320, using the expressions for the emission and absorption coefficients for gyrosynchrotron radiation from \citet{1984AuJPh..37..675R}. The resulting estimates of coronal magnetic field strength and electron density are highly degenerate with emission region size and electron power-law index, but are none the less consistent with those presented in \citet{2017MNRAS.465.1995M}. Future simultaneous observations with a low-frequency array (such as the upgraded GMRT, now with 300~MHz instantaneous bandwidth), and high-frequency array (e.g. the Karl G. Jansky Very Large Array or Australia Telescope Compact Array) will obtain instantaneous snapshots covering the optically thick and thin parts of the SED, yielding more accurate estimates of coronal properties responsible for the radio emission.


\section*{Acknowledgements}
\addcontentsline{toc}{section}{Acknowledgements}
We extend our thanks to Ian Stevens and Samuel George for informing us about their archival GMRT observations of UCDs re-analysed in this work. AZ thanks Huib Intema and Emil Lenc for helpful discussions regarding data reduction for this work. AZ acknowledges support from an Australian Government Research Training Program (RTP) Scholarship. TM acknowledges the support of the Australian Research Council through grant FT150100099. DLK was supported by NSF grant AST-1412421. PC acknowledges support from the Department of Science and Technology via a SwaranaJayanti Fellowship award (file no. DST/SJF/PSA-01/2014-15).
This research was conducted by the Australian Research Council Centre of Excellence for All-sky Astrophysics (CAASTRO), through project number CE110001020. We thank the staff of the GMRT that made these observations possible. GMRT is run by the National Centre for Radio Astrophysics of the Tata Institute of Fundamental Research. This research has made use of NASA's Astrophysics Data System Bibliographic Services. This research has made use of the SIMBAD database, operated at CDS, Strasbourg, France. This research has made use of the VizieR catalogue access tool, CDS, Strasbourg, France \citep{2000A&AS..143....9W}. The original description of the VizieR service was published in \citet{ 2000A&AS..143...23O}. This research made use of the following Python software packages: APLpy, an open-source plotting package for Python hosted at \url{http://aplpy.github.com}; Matplotlib \citep{2007CSE.....9...90H}; NumPy \citep{2011arXiv1102.1523V}; and Astropy, a community-developed core Python package for Astronomy \citep{2013A&A...558A..33A}.



\bibliographystyle{mnras}





\bsp	
\label{lastpage}
\end{document}